# Laser Transmission Studies with Magnetic Nanofluids

Chintamani Pai[1,2], H. Muthurajan[2], Nooris Momin[1], Radha S.[1] and R. Nagarajan[3]

[1] Department of Physics, University of Mumbai, Santacruz (E), Mumbai-400098, INDIA
[2] National Centre for Nanoscience and NanotechnologyUniversity of Mumbai, Santacruz (E), Mumbai-400098, INDIA
[3] UM-DAE Centre for Excellence in Basic Sciences, Santacruz (E), Mumbai-400098, INDIA

## Abstract:

Transmission of He-Ne (632 nm, 10 mW) Gaussian laser beam through Hexane and Water based magnetic nanofluids containing $Fe_3O_4$ nanoparticles show strong non-linear and magneto-optical effects. Application of external magnetic field (up to 1.7 Wb/m$^2$) perpendicular to the incident laser beam produces a change in forward scattered pattern of the incident laser beam. Dependence of forward scattered patterns in presence of external magnetic field has been studied. Image processing has been carried out to understand spatial distribution of the forward scattered patterns and temporal evolution of patterns involving particle image velocimetry technique. Change in non-linear refractive index is estimated for samples showing self-diffraction arising from higher order non-linear optical effect. Observed effects are useful for understanding light scattering from magnetic nanofluids and developing optofluidic devices and sensors.

## Introduction:

Magnetic fluids have emerged as smart nanofluids in the last few decades. Magnetic fluids were invented by NASA in 1960s to resolve fuel injection problems in microgravity. Since then, varieties of applications have been found by researchers. Magnetic particles dispersed in liquid have been used for advanced sensing applications in biomedical, mechanical and optical engineering by tuning the properties of magnetic nanoparticles and liquid media [Laskar 2012, Raj 2004, Vekas 2004, Scherer 2005, Kosea 2009, Rozhkova 2009, Kim 2010, Trana 2010].

Magnetic fluid is a stable colloidal suspension containing spherical magnetic particles of sizes ranging from 5 to 200 nm. The physical properties of the fluid are strongly influenced by the nature of size distribution of magnetic particles, the coating and the dispersion medium.

Studies of interaction of light with magnetic nanofluids have been conducted by various research groups [Laskar 2012]. External magnetic field induces optical anisotropy leading to magneto-optical effects. These include, optical birefringence, Faraday rotation, dichroism and different types of light scattering [Laskar 2012, Bacri 1982, Yusuf 2009, Rablau 2008, Philip 2008, Mohan 2012, Nair 2008, Nair 2008, Brojbasi 2015, Malynych 2010, Tokarev 2010].

In the absence of an external magnetic field, if the magnetic interaction energy between the nanoparticles in a dispersion medium is smaller than thermal energy $k_BT$, hence, no aggregation takes place and particles remain dispersed in the medium. When an external magnetic field is applied, magnetic moments orient along the external magnetic field and this magnetic energy contribution dominates the thermal energy. Ordered structures are formed due to this interaction. Chain formation of magnetic nanoparticles occurs along the magnetic field [Butter 2003, Nihad 1989, Philip 2012]. One of the well known applications of magnetic fluids in heat transfer technologies is attributed to aggregation under external field [Wang 2012]. Chain formation along the magnetic field also brings optical anisotropy in the medium [Nair 2008]. Interplay between the Brownian motion and magnetic interaction controls the scattering of the incident laser beam on magnetic nanofluid.

Light interaction with these ordered structures is interesting to study from both fundamental science and applications perspective. Light scattering from magnetic fluids has a strong correlation between aggregation kinetics of nanoparticles [Laskar 2010, Jones 1988]. The light scattering studies provide useful information about the clustering mechanism of nanoparticles [Hoffmann 2003].

Our earlier studies on laser beam transmission through Hexane based magnetic nanofluids show self diffraction patterns that is attributed to thermal lensing and non-linear optical effects [Radha 2014]. This study was further extended to water based magnetic nanoparticles with different organic coating like Polyacrylic acid (PAA). In this paper, we discuss our laser transmission experiments carried out with magnetic fluids involving comparative study of the effects in our samples. We estimate non-linear refractive index by analyzing self diffraction patterns and further image processing to analyze spatial distribution of transmitted laser beam.

## Material and methods:

All the samples of $Fe_3O_4$ nanoparticles were prepared by standard co-precipitation technique. The hydrodynamic radius of the $Fe_3O_4$ particles coated with Oleic acid and Polyacrylic acid as measured by dynamic light scattering (DLS) was 40 nm and 82 nm respectively.

The saturation magnetization measurements on a SQUID magnetometer (Quantum Design MPMS5) showed an $M_s$ of 40 emu/g and nearly zero coercivity at room temperature confirming the superparamagnetic nature. Laser transmission studies were carried out by passing He-Ne Gaussian laser beam (632 nm, 10 mW) through the dispersions of magnetic nanoparticles in Hexane and Water separately. External magnetic field was set perpendicular to the incident laser beam. Electromagnet used was capable of producing magnetic field 0 to 1.7 Wb/m$^2$ by interfacing it with PC using an in-house built software developed by the authors. In some cases, strong permanent magnet of identical strength was used to study light scattering. Forward scattered patterns were recorded using a CCD camera.

Open source software like VirtualDub and ImageJ were used for video and image processing. **Fig.1** shows the schematic of the experiment. VirtualDub was used to extract frames from a video file. Later images were processed using ImageJ using the modules for 3D intensity profiles, surface plot and particle image velocimetry (PIV) for cross-correlation analysis. The vector plot shows the arrows pointing movement due to changes in spatial distribution of transmitted light when any two frames are correlated.

## Results and discussions:

Hexane and Water based magnetic nanofluids show different response to the incident laser beam in presence of external magnetic field. Separate studies were carried out to investigate laser interaction with the samples.

We observe well known self-diffraction pattern of laser in magnetic nanofluids (first frame in **Fig. 3**). This pattern can be broadly understood as follows. When a laser beam is passed through a magnetic nanofluid, thermal lensing [Gordon 1965] and non-linear optical effects have been observed [Turek 1999, Haas 1975, Du 1994, Du 1995, Du 1998, Luo 1998, Luo 1999, Parekh 2005, Pu 2004]. Third order non-linear susceptibility of the medium produces effects like self-diffraction. Thermal and concentration diffusion of particles occur due to local heating by the incident laser beam inside magnetic nanofluid [Tabiryan 1998].

It is to be noted that the pattern is not a full circle but nearly semi-circular. This is attributed to the effect of gravity compressing the patterns in the vertical direction [Ji 2006]. This is verified in our samples by passing the laser beam at different depths [Radha 2014, Pai 2014].

The details of self-diffraction pattern are understood in terms of profile of the local index of refraction resulting in phase change $\delta\phi$ of the laser beam [Du 1998]. Change in non-linear refractive index $\delta n$ based on the number of rings N in the diffraction pattern is given by [Du 1998],

$$\delta n = \frac{N\lambda}{L} = n_2 I \qquad (1)$$

$$n = n_0 + \delta n \qquad (2)$$

Note that $\delta n$ is proportional to intensity, $I$, of the incident laser beam. Self-diffraction pattern carries information about the development of temperature profile and spatial distribution of the particles in dispersion medium. The variation of refractive index in general would arise from three contributions; the thermal gradient, concentration gradient and magnetic field. The details of each of the contributions would differ depending on the nature of the experiment.

**Observations under the magnetic field:**

Application of magnetic field to our samples of Hexane based magnetic nanofluid and observation under inverted microscope shows typical chain formation of magnetic nanoparticles. Micron size chains are formed along the direction of the external magnetic field (**Fig. 2**).

*A. Forward scattered patterns ($Fe_3O_4$ in Water 10 mg/ml)*

In case of water based magnetic fluid, a linear streak is observed when magnetic field is applied perpendicular to the incident laser beam. **Fig. 3** shows the forward patterns with and without magnetic field at

0.170 Wb/m$^2$. Identical pattern was observed at very low magnetic field 0.005 Wb/m$^2$ but it takes finite time (~ 120 sec) to develop the stable linear streak with lobes. **Fig. 4** shows the development of the patterns using 3D surface plot at 0.005 Wb/m$^2$. It clearly reveals the enhancement in the intensity of lobes with time. Intensity of lobes at 0.005 Wb/m$^2$ is lower than the lobes at 0.170 Wb/m$^2$. Because at higher field fully grown chains undergo zippering transitions causing enhanced intensity than at lower field [Brojabasi 2015]. The temporal scattered patterns seem to arise from resultant of multiple diffraction patterns due to individual chains of magnetic nanoparticles behaving like cylindrical shaped particles in external magnetic field [Laskar 2012].

*B. Rotation of patterns (Fe$_3$O$_4$ in Hexane 30 mg/ml)*

Self diffraction patterns with laser beam incident at an intermediate depth on cuvette containing magnetic fluid, show distinct rotation as magnetic field is increased (**Fig. 5**). The change in rotation angle with increase in magnetic field is given in **Table 1.** Change in non-linear refractive index varies from 1.896 x 10$^{-3}$ to 2.528 x 10$^{-3}$. The field dependent distortion has been observed by others also and has been attributed to Kelvin Body force [Luo 1999]. This destabilizes the earlier developed temperature profile and spatial distribution of the particles. Self diffraction patterns tend to show distortion due to such convective instabilities interfering with refractive index profile in the magnetic nanofluid.

*C. Transient analysis (Fe$_3$O$_4$ in Hexane 30 mg/ml)*

We also observed dynamics of the formation of the self diffraction pattern by studying the transient response of the magnetic fluid to laser field in the presence of magnetic field. Time evolution of the laser transmission pattern was studied in presence of the magnetic field of 1.7 Wb/m$^2$. **Fig. 6** shows consecutive extracted frames showing the formation of self diffraction patterns at 70 ms scale. As the shutter opens, the bright spot of the transmitted laser beam starts turning into nearly circular pattern with 4-5 rings. Effect of gravity resulting into vertical suppression of patterns is not observed [Ji 2006,]. **Table 2** gives the estimation of change in non-linear refractive index based on the number of rings in the patterns. 3D intensity profile of the final stable pattern is shown in **Fig. 7**. The change in non-linear refractive index is estimated for hexane based magnetic nanofluids using self diffraction patterns. It is found to be of the order 10$^{-3}$, clearly indicating contribution from thermal lensing phenomenon.

*D. Use of PIV technique for temporal analysis of patterns*

PIV technique is used for cross-correlating self diffraction patterns to understand how intensity variations occur at millisecond scale. **Fig. 8** shows the vector field plot of forward scattered patterns for case B and case C. **Fig. 8b** shows clear divergence in the patterns evolving at 70 ms scale as refractive index gradient develops. **Fig. 8a** does not show any such divergence but rotational appearance. This is useful in understanding the evolution of patterns arising because of thermal and concentration diffusion phenomena as refractive index gradient is being established.

## Conclusion:

Laser transmission studies through magnetic nanofluids show a variety of magneto-optical effects that depend on the magnetic field as well as physical properties of the nanoparticles and dispersion media. This is useful for studying laser nanofluid interaction and resulting dynamic forward scattering. Image processing helps to understand spatial distribution of forward scattered patterns. Dynamical changes in intensity are analyzed using PIV technique. Utilization of the observed effects is promising for developing magnetic fluid based sensors and optofluidic devices.

## Acknowledgement:

We extend our sincere thanks to Prof Anuradha Misra (Head, Dept of Physics, University of Mumbai), Prof R.V. Hosur (Director, UM-DAE CBS) for support and Prof. A. K. Nigam (TIFR, Mumbai) for magnetic measurements.

## References:

Bacri JC, Salin D (1982), "Optical scattering on ferrofluid aggregates", J. Physique-Lett. 43, L-771.
Butter, Bomans PHH, Frederik PM, Vroege GJ, Phillipse AP (2003), "Direct observation of dipolar chains in iron ferrofluids by cryogenic electron microscopy", Nat. Mat. vol. 2, p. 88.
Brojabasi S, Muthukumaran T, Laskar JM, John Philip (2015), "The effect of suspended Fe$_3$O$_4$ nanoparticle size on magneto-optical properties of ferrofluids", Opt. Comm., vol. 336, pp. 278–285.

Du T, Luo W (1995), "Dynamic Interference Patterns from Ferrofluids", Mod. Phys. Lett. B, vol. 9, pp. 1643-1647.

Du T, Luo W (1998), "Nonlinear optical effects in ferrofluids induced by temperature and concentration cross coupling", vol. 72, pp. 272-274.

Du T, Yuan S, Luo W (1994), "Thermal lens coupled magneto-optical effect in a ferrofluid", *Appl. Phys. Lett.,* vol. 65, pp. 1844-1846.

Gordon JP, Leite RCC, Moore RS, Porto SPS, Whinnery JR (1965), "Long transient effects in lasers with inserted liquid samples", J. App. Phys., vol. 36, p.3.

Haas WL., Adams JE (1975), "Diffraction effects in ferrofluids", App. Phys. Lett., vol. 27, p. 571.

Hoffmann B, Koehler W (2003), "Reversible light-induced cluster formation of magnetic colloids", J. Mag. Magn. Mat., vol. 262, pp. 289–293.

Ji W, Chen W, Lim S, Lin J, Guo Z (2006), "Gravitation-dependent, thermally-induced self-diffraction in carbon nanotube solutions", Opt. Exp., vol. 14, pp. 8958-8966.

Jones GA, Niedoba H (1988), "Filed induced agglomeration in thin films of aqueous based magnetic fluids", J. Mag. Magn. Mat., vol. 73, pp. 33-38.

Kim D, Rozhkova EA, Ulasov IV, Bader SD, Rajh T, Lesniak MS, Novosad V (2010), "Biofunctionalized magnetic-vortex microdiscs for targeted cancer-cell destruction", vol. 9, pp. 165–171.

Kosea AR., Fischerb B, Maoc L, Kosera H (2009), "Label-free cellular manipulation and sorting via biocompatible ferrofluids", PNAS, vol. 106, 21478–21483.

Laskar J, Broajbasi S, Raj B, Philip J, (2012) "Comparison of light scattering from self assembled array of nanoparticles chains with cylinders", vol. 285, pp. 1242-1247.

Laskar J, Philip J (2012), "Optical properties and applications of ferrofluids— A review", J. Nanofluids vol. 1, pp. 3-20.

Laskar J M, Philip J, Raj B (2010), "Experimental investigation of magnetic-field-induced aggregation kinetics in non-aqueous ferrofluids", Phys. Rev. E., vol. 82, p. 021402.

Luo W, Du T, Huang J (1998), "Field Induced Instabilites in a Magnetic Fluid, J. Mag. Magn. Mat., vol. 201, pp. 88-90.

Luo W, Du T (1999), "Novel convective instabilities in a magnetic fluid", Phys. Rev. Lett., vol. 82, 4134-4137.

Malynych S Z, Tokarev A, Hudson S, Chumanov G, John B, Kornev KG (2010), "Magneto-controlled illumination with opto-fluidics", J. Mag Magn Mat., pp. 1894–1897.

Mohan S, Sharma D, Deshpande AA, Mathur D, Ramachandran H, Kumar N, (2012), "Light scattering from a magnetically tunable dense random medium with dissipation: ferrofluid", Euro. Phys. J. D, vol. 66, p. 30.

Nair S, Thomas J, Sandeep C S S, Anantharaman M R, Philip Reji (2008), " An optical limiter based on ferrofluids", App. Phys. Lett., vol. 92, p. 171908.

Nair S, Xavier F, Joy P A, Kulkarni S D, Anantharaman M R (2008), J. Mag. Magn. Mat., vol. 320, pp. 815-820.

Nihad A Yusuf (1989), "Field and concentration dependence of chain formation in magnetic fluids", J. Phys. D: Appl. Phys., vol. 22, pp. 1916-1919.

Parekh K, Patel R, Upadhyay RV, Mehta RV (2005), "Field-induced diffraction patterns in a magneto-rheological suspension", J. Mag. Magn. Mat. Vol. 289, pp. 311-313.

Pai C, Mohan S, Radha S (2014), " Transient optical phenomenon in ferrofluids", Proc. Engg., vol. 76, pp. 74-79.

Philip J, Laskar JM, Raj B (2008)," Magnetic Field induced extinction of light in a suspension of $Fe_3O_4$ nanoparticles", App. Phys. Lett., vol. 92, 221911.

Pu S, Chen X, Liao W, Chen L, Chen Y, Xia Y (2004), "Laser self-induced thermo-optical effects in a magnetic fluid", J. Appl. Phys., vol. 96 pp. 5930-5932.

Rablau C, Vaishnava P, Sudakar C, Tackett R, Lawes G, Naik R (2008), "Magnetic field induced optical anisotropy in ferrofluids: A time dependent light scatterring investigation", Phys. Rev., E, vol. 78, 051502.

Radha S, Mohan S, Pai C (2014)," Diffraction patterns in ferrofluids: Effect of magnetic field and gravity", Phys. B, vol. 448, 341-345.

Raj K, Moskowitz B, Tsuda S (2004), "New commercial trends of nanostructured ferrofluids", Ind. J. Engg. and Mat. Sci.,vol.11, pp. 253-261.

Rozhkova EA, Novosad V, Kim DH, Pearson J, Divan R, Rajh T, Bader S D (2009), "Ferromagnetic microdisks as carriers for biomedical applications", J. App. Phys., vol. 105, p. 07B306.

Scherer C, Figueiredo Neto AM (2005), "Ferrofluids: properties and applications", Brazilian J. Phys., vol. 35, 3A.

Tabiryan NV, Luo W (1998), " Soret feedback in thermal diffusions of suspensions", Phys. Rev. E., vol. 57, pp. 4431-4440.

Tokarev A, Rubin B, Bedford M, Kornev K G (2010), "Magnetic nanorods for optofluidic applications" , 8Th International conference on the scientific and clinical applications of magnetic carriers.

Turek I, Stelina J, Musil C, Timko M, Kopcansky P, Koneracka M, Tomco L (1999) , "The effect of self diffraction in magnetic fluids", J. of Magn. Mag. Mat., vol. 201, pp. 167-169.

Trana N, Webster TJ (2010), "Magnetic nanoparticles: biomedical applications and challenges", J. Mater. Chem., vol. 20, pp. 8760-8767.
Vekas L (2004), "Magnetic nanofluids properties and some applications", Rom. J. Phys., vol. 49, pp. 707-721.
Wang J J, Zheng R T, Gao J W, Chen G (2012), "Heat conduction mechanisms in nanofluids and suspensions", vol. 7, pp. 124-136.
Yusuf NA and Abu-Aljarayesh IO (2009), "Magneto-optical and magneto-dielectric anisotropy effects in magnetic fluids", Jordon J. Phys., vol. 22, pp. 1-46.

**Figures:**

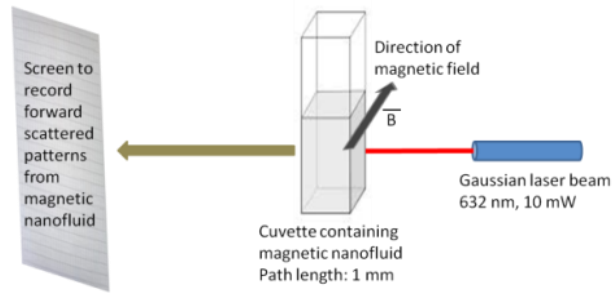

Fig. 1. Schematic of the experimental set-up

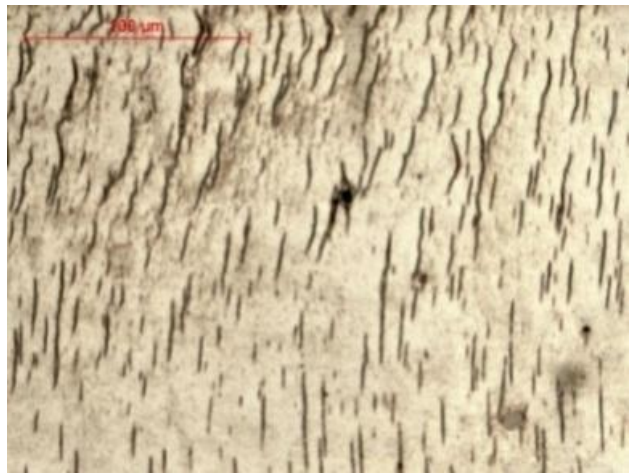

Fig. 2. Chains of magnetic nanoparticles along the magnetic field

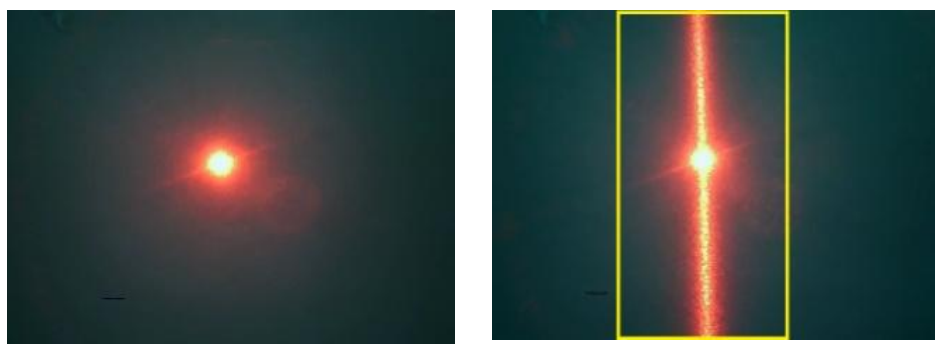

Zero magnetic Field                  0.17Wb/m$^2$

Fig.3. Forward scattered patterns in presence of magnetic field [$Fe_3O_4$ nanoparticles (10 nm) in Water (10mg/ml)].

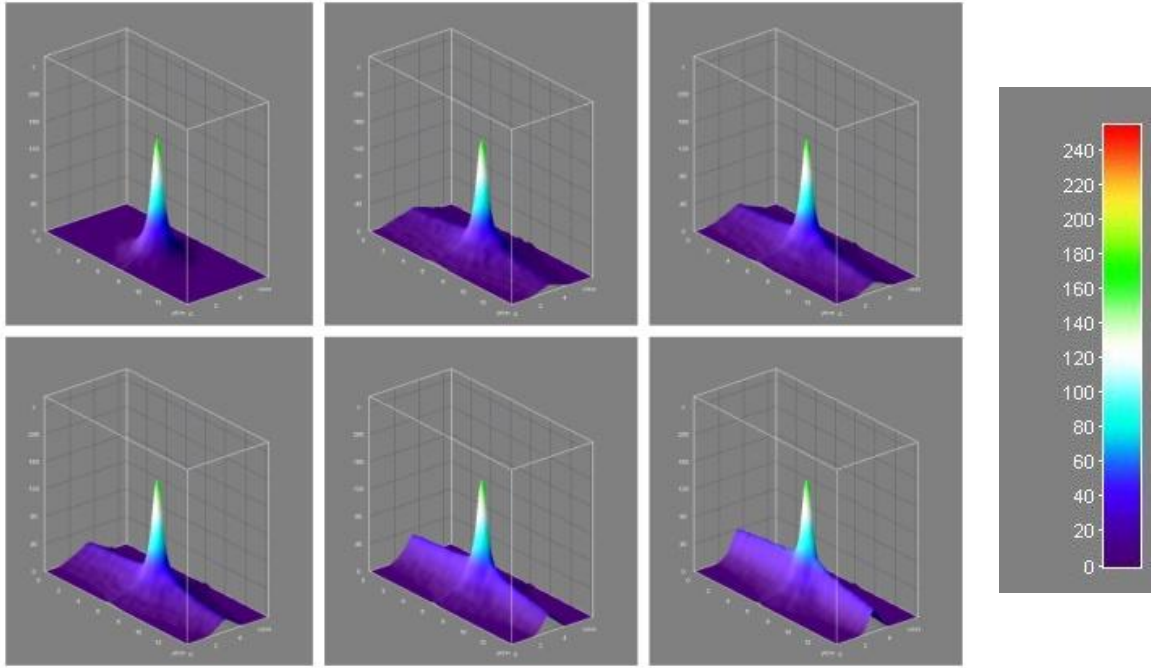

Fig.4. Development of forward scattered patterns in at 0.005 Wb/m$^2$.
Fe$_3$O$_4$ nanoparticles (10 nm) in Water (10mg/ml)].

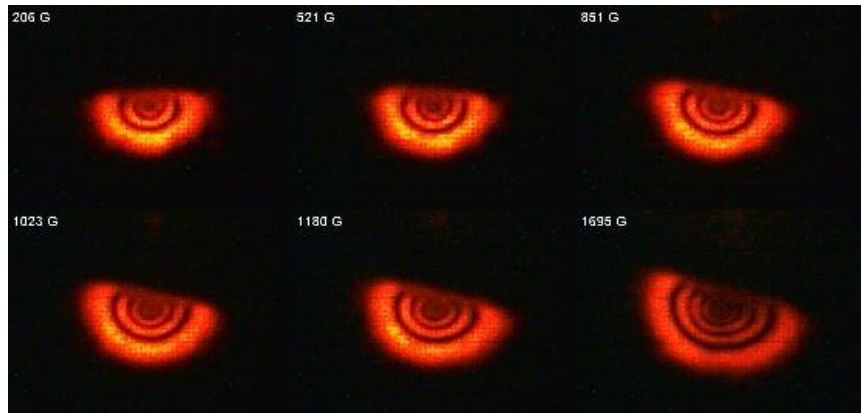

Fig. 5. Patterns showing gradual rotation with increase in magnetic field field
[Fe$_3$O$_4$ nanoparticles (40 nm) in Hexane (30mg/ml)].

| Magnetic Field (10$^{-4}$ Wb/m$^2$) | Rotation (Deg) |
|---|---|
| 206 | 0 |
| 521 | 5.42 |
| 851 | 13.83 |
| 1023 | 14.31 |
| 1180 | 15.19 |
| 1695 | 16.14 |

Table 1. Rotation of self diffraction patterns with magnetic field

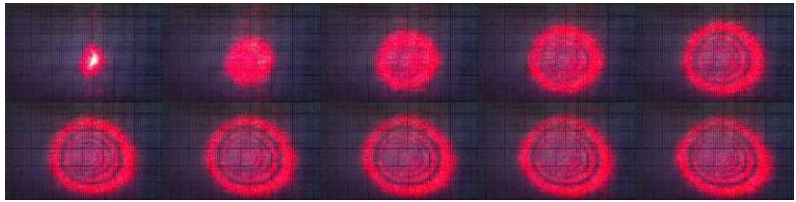

Fig. 6. Time evolution in magnetic field 0.17Wb/m$^2$
[Fe3O4 nanoparticles (40 nm) in Hexane (30mg/ml)].

| Time (ms) | No. of rings, N | Change in non-linear refractive index, $\Delta n_2$ (x $10^{-3}$) |
| --- | --- | --- |
| 210 | 2 | 1.264 |
| 280 | 4 | 2.528 |
| 350 | 5 | 3.160 |
| 420 | 6 | 3.792 |

Table 2. Calculation of change in non-linear refractive index as self diffraction patterns develop in magnetic field 0.17 Wb/m2 [Fe$_3$O$_4$ nanoparticles (40 nm) in Hexane (30mg/ml)].

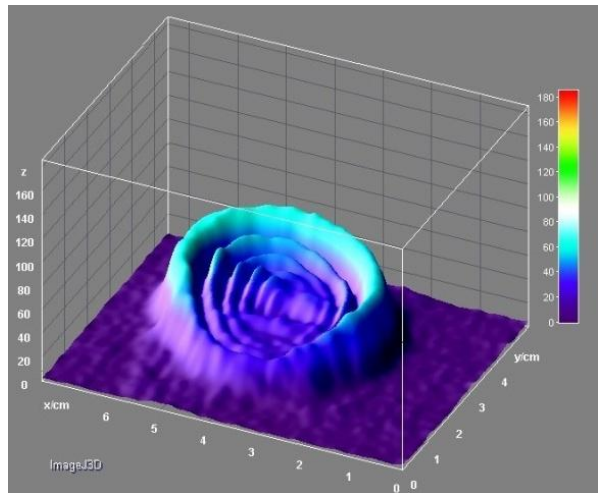

Fig. 7. 3D intensity profile of the final stable pattern in magnetic field
[Fe$_3$O$_4$ nanoparticles (40 nm) in Hexane (30mg/ml)].

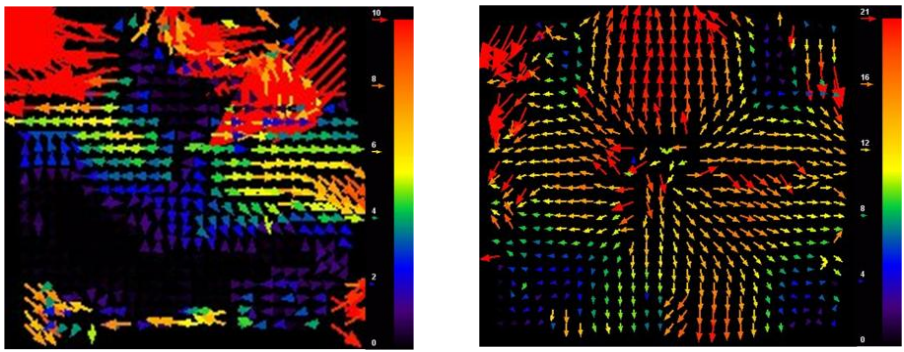

Fig. 8 Vector field of selected a) patterns showing rotation b) consecutive patterns using PIV analysis
[Fe$_3$O$_4$ nanoparticles (40 nm) in Hexane (30mg/ml)].